\documentclass[letter]{jpsj2}

\def\e{\mathrm{e}}
\def\i{\mathrm{i}}
\def\Define{:=}

\def\nn{\nonumber}

\def\sgn{\mathrm{sgn}}
\renewcommand{\leq}{\leqslant}

\title{Resonance in an open quantum dot system 
with a Coulomb interaction: \\ a Bethe-ansatz approach}

\author{Akinori \textsc{Nishino}%
\thanks{E-mail address: nishino@iis.u-tokyo.ac.jp} 
and Naomichi \textsc{Hatano}
\thanks{E-mail address: hatano@iis.u-tokyo.ac.jp}}

\inst{Institute of Industrial Science, The University of Tokyo,
4--6--1 Komaba, Meguro-ku, Tokyo, 153--8505}

\abst{An open quantum system consisting
of a quantum dot with a Coulomb interaction
and two leads without interactions is studied.
The many-body scattering states are constructed
with the Bethe-ansatz approach.
The expectation value of the electric current 
is exactly calculated for the scattering states to observe
resonance peaks due to many-body scattering.}

\kword{scattering theory, open quantum systems, resonance,
interacting resonant-level model, 
quantum dots, Bethe ansatz}

\begin{document}
\maketitle

The purpose of this letter is to observe resonance
in an open quantum system with a Coulomb interaction.
The system that we study is the two-lead interacting 
resonant-level model (IRLM), which consists of
two leads of non-interacting electrons
that interact with an electron on a quantum dot
in between the two leads.
We obtain $N$-electron scattering states
for arbitrary $N$, generalizing
the Bethe-ansatz approach to open systems.
By using the scattering states, we exactly calculate
the quantum-mechanical expectation value of the electric current 
through the quantum dot, thereby observing resonance peaks.
Some of the resonance peaks appear 
only when the interaction exists;
they reflect the effect of many-body scattering.

The resonance of many-body scattering
that we observe in the quantum-mechanical 
expectation value has not been found in previous works with
the Bethe ansatz.
The Bethe-ansatz approach has provided a nonperturbative 
method of studying equilibrium states
of interacting quantum systems
including the Kondo problem~\cite{Andrei_80PRL,Wiegmann_80PLA,%
Filyov-Wiegmann_80PLA,Andrei-Furuya-Lowenstein_83RMP,%
Tsvelick-Wiegmann_83AP}.
The approach is now used to discuss transport properties of 
mesoscopic systems.
Konik {\it et al.}~\cite{Konik-Saleur-Ludwig_01PRL,%
Konik-Saleur-Ludwig_02PRB} studied
transport properties of the Anderson model
in the thermodynamic limit of a closed system 
with {\it periodic} boundary conditions.
Our scattering states, in contrast,
appear only in open systems;
they are constructed 
{\it without} imposing periodic boundary conditions.
By extending the Bethe-ansatz approach,
Mehta and Andrei \cite{Mehta-Andrei_06PRL} 
studied the two-lead IRLM as an open system
to obtain $N$-electron scattering states giving 
nonequilibrium steady states in the limit $N\to\infty$.
In their study, however,
the quantum-mechanical expectation value of the current 
does not depend on the interaction;
the effect of the interaction appears only
in the statistical-mechanical expectation value
as modification of the Fermi distribution in the leads. 
Thus our results are different from
the previous ones.

There has recently been a great deal of interest 
in mesoscopic systems with interacting electrons.
Experiments suggest that interactions 
are essential in understanding their transport properties~\cite{%
Ralph-Buhrman_92PRL,Ralph-Buhrman_94PRL,%
GoldhaberGordon_98Nature,GoldhaberGordon_98PRL,%
Cronenwett_98Science}.
The perturbation theory tells us that
the effect of interactions is observed as resonance peaks
of the electrical conductance~\cite{Meir-Wingreen-Lee_91PRL,%
Yeyati-MartinRodero-Flores_93PRL}.
For non-interacting open quantum systems,
the relation between quantum mechanical scattering states
and nonequilibrium steady states is well 
investigated~\cite{Datta,Zagoskin}.
However, the relation in interacting open quantum systems 
has not been clarified, excepting Schiller and Hershfield's 
result \cite{Schiller-Hershfield}
at a special point of the interaction parameter where
an interacting system is mapped to a non-interacting one.
The present study gives a steady step toward 
an exact analysis of interacting open quantum systems 
out of equilibrium.

The Hamiltonian of the two-lead IRLM is given by
\begin{align}
\label{eq:IRLM_12}
H\!=\!
&\sum_{i=1,2}\hspace{-3pt}\Big(\!
  \int_{-\frac{L}{2}}^{\frac{L}{2}}
  \hspace{-5pt}dz\hspace{1pt}
  c^{\dagger}_{i}(z)\frac{1}{\i}\frac{d}{dz}c_{i}(z)
 +\frac{t}{\sqrt{2}}
  \big(c^{\dagger}_{i}(0)d\!+\!d^{\dagger}c_{i}(0)\big)
  \Big) \nn\\
&+\epsilon_{d}d^{\dagger}d
 +U\sum_{i=1,2}c^{\dagger}_{i}(0)c_{i}(0)d^{\dagger}d,
\end{align}
where $t(>0)$ is the transfer integral between each lead 
and the quantum dot, $\epsilon_{d}$ is the gate energy of the dot 
and $U(>0)$ expresses the Coulomb repulsion.
The dispersion relation in the leads is linearized
in the vicinity of the Fermi energy to be $E=k$, under
the assumption that $t$, $\epsilon_{d}$ 
and $U$ are small compared with the Fermi 
energy~\cite{Andrei_80PRL,Wiegmann_80PLA,Filyov-Wiegmann_80PLA}.
We stress that
we treat the system as an open system in the limit $L\to\infty$.
The one-lead IRLM with periodic boundary conditions
was studied with the Bethe ansatz~\cite{Filyov-Wiegmann_80PLA}.
Our purpose is to investigate, for scattering states,
the electric current through the quantum dot,
\begin{align}
\label{eq:current-op}
I\!\Define\!\Dot{N_{2}}\!-\!\Dot{N_{1}}
 \!=\!\frac{t}{\sqrt{2}\i}\sum_{i=1,2}(-)^{i} 
      \big(c_{i}^{\dagger}(0)d-d^{\dagger}c_{i}(0)\big).
\end{align}

We derive the Schr\"odinger equations for the system.
After the transformation
$c_{1/2}(z)=\big(c_{e}(z)\pm c_{o}(z)\big)/\sqrt{2}$,
the Hamiltonian~\eqref{eq:IRLM_12} is decomposed into
the even and odd parts~\cite{Mehta-Andrei_06PRL}. 
Due to the relations $[H, N_{e}+N_{d}]=[H, N_{o}]=0$
for the number operators 
$N_{e/o}=\int dz c_{e/o}^{\dagger}(z)c_{e/o}(z)$ and
$N_{d}=d^{\dagger}d$,
the set $(N_{e}+N_{d},N_{o})$
gives a good quantum number.
The $N$-electron state in the subspace with 
$N_{o}=n, (0\leq n\leq N)$
is generally expressed in the form
\begin{align}
\label{eq:general-N-state}
|\psi\rangle
&=\Big(\!\int\!\!dz\,
  g^{(n)}(z)
  c^{\dagger}_{e}(z_{1})\cdots c^{\dagger}_{e}(z_{N-n})
  c^{\dagger}_{o}(z_{N-n+1})\cdots c^{\dagger}_{o}(z_{N})
  \nn\\
&\quad+\!\int\!\!dz\,
  e^{(n)}(z)
  c^{\dagger}_{e}(z_{1})\cdots 
  c^{\dagger}_{e}(z_{N-n-1})
  d^{\dagger}
  c^{\dagger}_{o}(z_{N-n})\cdots 
  c^{\dagger}_{o}(z_{N-1})
  \Big)|0\rangle,
\end{align}
where $g^{(n)}(z)=g^{(n)}(z_{1},\ldots,z_{N})$ 
for $0\leq n\leq N$
and $e^{(n)}(z)=e^{(n)}(z_{1},\ldots,z_{N-1})$
for $0\leq n\leq N-1$ are functions to be determined.
We also set $e^{(N)}(z)=0$ for convenience.
The eigenvalue problem $H|\psi\rangle=E|\psi\rangle$
is cast into a set of the Schr\"odinger equations
\begin{align}
\label{eq:Sch-eq_N-state_IRLM}
&(N-n)\Big(
 \sum_{i=1}^{N}
 \frac{1}{\i}\frac{\partial}{\partial z_{i}}\!-\!E
 \Big)g^{(n)}(z)
 +t\sum_{i=1}^{N-n}(-)^{N-n-i}\delta(z_{i})
 e^{(n)}(\ldots,z_{i-1},z_{i+1},\ldots)
 =0,
 \nn\\
&\Big(\sum_{i=1}^{N}
 \frac{1}{\i}\frac{\partial}{\partial z_{i}}\!-\!E
 \Big)
 g^{(N)}(z)
 =0,
 \nn\\
&\Big(\sum_{i=1}^{N-1}
 \Big(\frac{1}{\i}\frac{\partial}{\partial z_{i}}
 \!+\!U\delta(z_{i})\Big)\!+\!\epsilon_{d}\!-\!E\Big)
 e^{(n)}(z) 
 \!+\!t(N\!-\!n)
 g^{(n)}(\ldots,z_{N-n-1},0,z_{N-n},\ldots)=0,
\end{align}
where $0\leq n\leq N\!-\!1$.
In what follows, we use the variables $x_{i}$ and $y_{i}$
to express the coordinates of the leads $e$ and $o$, 
respectively: 
$g^{(n)}(z)=g^{(n)}(x_{1},\ldots,x_{N-n},y_{1},\ldots,y_{n})$ 
and $e^{(n)}(z)
=e^{(n)}(x_{1},\ldots,x_{N-n-1},y_{1},\ldots,y_{n})$.
The set of eigenfunctions in the one-electron sector
with $E=k$ is given by
\begin{align}
&g^{(0)}(x_{1})=g_{k}(x_{1})
 \Define\frac{2\e^{\i kx_{1}}}{1+\e^{\i\delta_{k}}}
 \big(\theta(-x_{1})+\e^{\i\delta_{k}}\theta(x_{1})\big), \nn\\
&e^{(0)}=e_{k}
 \Define\frac{t}{k-\epsilon_{d}},
 \nn\\
&g^{(1)}(y_{1})=h_{k}(y_{1})
 \Define\frac{2\e^{\i ky_{1}}}{1+\e^{\i\delta_{k}}} \nn
\end{align}
with the phase shift $\delta_{k}\Define
-2\arctan\big(t^{2}/2(k-\epsilon_{d}))$
of one-body scattering at $x_{1}=0$ in the lead $e$
and the step function $\theta(z)$.
Note that the eigenfunction $g_{k}(x_{1})$ 
is discontinuous at $x_{1}=0$.

We construct an $N$-electron eigenstate
with the Bethe ansatz.
It is different from 
the one obtained by Mehta and Andrei~\cite{Mehta-Andrei_06PRL}.
To demonstrate the difference, 
we first consider the case $N=2$.
The set of two-electron eigenfunctions 
with the energy eigenvalue $E=k_{1}+k_{2}$ 
is assumed to be
\begin{align}
\label{eq:2-Bethe-eigenfunction}
&2g^{(0)}_{k_{1}k_{2}}(x_{1},x_{2})
 =Z_{k_{1}k_{2}}(x_{1}-x_{2})g_{k_{1}}(x_{1})g_{k_{2}}(x_{2})
 -Z_{k_{1}k_{2}}(x_{2}-x_{1})g_{k_{2}}(x_{1})g_{k_{1}}(x_{2}), 
 \nn\\
&e^{(0)}_{k_{1}k_{2}}(x_{1})
 =Z_{k_{1}k_{2}}(x_{1})g_{k_{1}}(x_{1})e_{k_{2}}
 \!-\!Z_{k_{1}k_{2}}(-x_{1})g_{k_{2}}(x_{1})e_{k_{1}}, 
 \nn\\
&g^{(1)}_{k_{1}k_{2}}(x_{1},y_{1})
 =X(x_{1}-y_{1})g_{k_{1}}(x_{1})h_{k_{2}}(y_{1}), 
 \nn\\
&e^{(1)}_{k_{1}k_{2}}(y_{1})=X(-y_{1})e_{k_{1}}h_{k_{2}}(y_{1}),
 \nn\\
&2g^{(2)}_{k_{1}k_{2}}(y_{1},y_{2})
 =h_{k_{1}}(y_{1})h_{k_{2}}(y_{2})
 \!-\!h_{k_{2}}(y_{1})h_{k_{1}}(y_{2}),
\end{align}
where the amplitudes
$Z_{k_{1}k_{2}}(z)$ and $X(z)$ are defined by
\begin{align}
&Z_{k_{1}k_{2}}(z)\Define
 \e^{-\frac{\i}{2}\varphi_{k_{1}k_{2}}}\theta(-z)
 +\e^{\frac{\i}{2}\varphi_{k_{1}k_{2}}}\theta(z),
 \nn\\
&X(z)\Define
 \e^{\frac{\i}{2}\eta}\theta(-z)
 +\e^{-\frac{\i}{2}\eta}\theta(z) \nn
\end{align}
with the phase shifts
$\varphi_{k_{1}k_{2}}
 \Define 2\arctan\big(\!-\!\frac{U}{2}
  \frac{k_{1}-k_{2}}{k_{1}+k_{2}-2\epsilon_{d}}\big)$ and
$\eta\Define 2\arctan(-U/2)$
of two-body scattering.
The eigenfunctions $g^{(0)}_{k_{1}k_{2}}(x_{1},x_{2})$ 
and $e^{(0)}_{k_{1}k_{2}}(x_{1})$ are the same Bethe eigenfunctions 
as those assumed in the one-lead IRLM \cite{Filyov-Wiegmann_80PLA},
although we do not impose periodic boundary conditions.
The eigenfunctions 
$g^{(1)}_{k_{1}k_{2}}(x_{1},y_{1})$ and 
$e^{(1)}_{k_{1}k_{2}}(y_{1})$
are obtained with separation of variables;
if we set $g^{(1)}_{k_{1}k_{2}}(x_{1},y_{1})=X(x_{1}-y_{1})
\Tilde{g}^{(1)}_{k_{1}k_{2}}(x_{1},y_{1})$
and $e^{(1)}_{k_{1}k_{2}}(y_{1})=X(-y_{1})
\Tilde{e}^{(1)}_{k_{1}k_{2}}(y_{1})$,
Eqs.~\eqref{eq:Sch-eq_N-state_IRLM}
are decoupled into even and odd parts,
and the eigenfunctions $\Tilde{g}^{(1)}_{k_{1}k_{2}}(x_{1},y_{1})$ 
and $\Tilde{e}^{(1)}_{k_{1}k_{2}}(y_{1})$ are given by the product
of eigenfunctions of the even and the odd parts.
The eigenfunctions $g^{(2)}_{k_{1}k_{2}}(y_{1},y_{2})$ 
should be a free fermion eigenfunction because of 
Eq.~\eqref{eq:Sch-eq_N-state_IRLM} for $g^{(N)}(z)$.
The phase shifts in our solution are different
for each two-body scattering in the lead $e$, in the lead $o$ 
and between the leads; this gives the resonance of
many-body scattering in the current expectation value,
as we shall see below.
In Mehta and Andrei's solution~\cite{Mehta-Andrei_06PRL},
on the other hand, the same phase shift $\varphi_{k_{1}k_{2}}$ 
of two-body scattering was adopted for all the two-electron 
eigenfunctions $g^{(n)}(z)$ and $e^{(n)}(z)$.

By exchanging $k_{1}$ and $k_{2}$ 
in both eigenfunctions $g^{(1)}_{k_{1}k_{2}}(x_{1},y_{1})$ 
and $e^{(1)}_{k_{1}k_{2}}(y_{1})$, 
we have another set of eigenfunctions 
$g^{(1)}_{k_{2}k_{1}}(x_{1},y_{1})$ 
and $e^{(1)}_{k_{2}k_{1}}(y_{1})$
with the same eigenvalue $E=k_{1}+k_{2}$.
In the limit $t, U\to 0$, 
the set $\{g^{(0)}_{k_{1}k_{2}}(x_{1},x_{2}),
g^{(1)}_{k_{1}k_{2}}(x_{1},y_{1}),
g^{(1)}_{k_{2}k_{1}}(x_{1},y_{1}),
g^{(2)}_{k_{1}k_{2}}(y_{1},y_{2})\}$ reproduces
a complete orthogonal system of two free fermions 
in the two leads, while Mehta and Andrei's 
solution~\cite{Mehta-Andrei_06PRL} does not.
In this sense,
our solution~\eqref{eq:2-Bethe-eigenfunction}
is more plausible than theirs.

In a way similar to the case $N=2$, we obtain a set of
$N$-electron eigenfunctions with the energy eigenvalue
$E=\sum_{i=1}^{N}k_{i}$ in the form
\begin{align}
\label{eq:N-Bethe-eigenfunction}
&g^{(n)}_{k}(x;y)
 =\frac{1}{(N\!-\!n)!n!}\!\!
  \sum_{P\in\mathfrak{S}_{N-n}\atop Q\in\mathfrak{S}_{n}}\!\!
  \sgn(PQ)Z^{(n)}_{k_{P}}(x;y)
  \prod_{i=1}^{N-n}g_{k_{P_{i}}}(x_{i})
  \prod_{j=1}^{n}h_{k_{N-n+Q_{j}}}(y_{j}),
  \nn\\
&e^{(n)}_{k}(x;y)
 =\frac{1}{(N\!-\!n\!-\!1)!n!}\!\!
  \sum_{P\in\mathfrak{S}_{N-n}\atop Q\in\mathfrak{S}_{n}}\!\!
  \!\!\!
  \sgn(PQ)\Tilde{Z}^{(n)}_{k_{P}}(x;y)
  \!\!\prod_{i=1}^{N-n-1}\!\!g_{k_{P_{i}}}(x_{i})
  e_{k_{P_{N-n}}}
  \prod_{j=1}^{n}h_{k_{N-n+Q_{j}}}(y_{j}),
\end{align}
where $\mathfrak{S}_{m}$ is the symmetric group
acting on the set $\{1,2,\ldots,m\}$ and
\begin{align}
&Z^{(n)}_{k_{P}}(x;y)
=\hspace{-8pt}\prod_{1\leq i<j\leq N\!-\!n}\hspace{-8pt}
 Z_{k_{P_{i}}k_{P_{j}}}(x_{i}-x_{j})
 \hspace{-5pt}\prod_{1\leq i\leq N\!-\!n\atop 1\leq j\leq n}
 \hspace{-5pt}
 X(x_{i}-y_{j}),
 \nn\\
&\Tilde{Z}^{(n)}_{k_{P}}(x;y)
=Z^{(n)}_{k_{P}}(x;y)\big|_{x_{N\!-\!n}=0}.
 \nn
\end{align}
Note that the amplitude $Z^{(n)}_{k_{P}}(x;y)$
and $\Tilde{Z}^{(n)}_{k_{P}}(x;y)$ are
given by the product of the amplitudes
of the two-electron eigenfunctions.
The $N$-electron eigenfunctions~\eqref{eq:N-Bethe-eigenfunction}
are indexed by a set of momenta $k=\{k_{1},\ldots,k_{N}\}$. 
We denote by $|k;n\rangle$ the eigenstate obtained by putting
the eigenfunctions $g^{(n)}_{k}(x;y)$ and $e^{(n)}_{k}(x;y)$
into \eqref{eq:general-N-state},
and call it a Bethe eigenstate.

We show that, 
for a fixed set of $N$ momenta $k=\{k_{1},\ldots,k_{N}\}$,
there exist $2^{N}$ degenerate Bethe eigenstates
with the energy eigenvalue $E=\sum_{i=1}^{N}k_{i}$.
For a fixed $n$, we consider $_{N}C_{n}$
ways of dividing the set $k$ 
into two subsets wherein the first subset contains $N-n$ elements
and the second subset contains $n$ elements.
It is convenient to index each way of dividing
by $k_{R}=\{k_{R_{1}},\ldots,k_{R_{N}}\}$ with
an element $R$ of the symmetric group 
$\mathfrak{S}_{N}$ satisfying $R_{1}<R_{2}<\cdots<R_{N-n}$
and $R_{N-n+1}<R_{N-n+2}<\cdots<R_{N}$.
The element $R$ is an element of
$\mathfrak{S}_{N}/(\mathfrak{S}_{N-n}\times\mathfrak{S}_{n})$,
where $\mathfrak{S}_{N-n}$ is the symmetric group acting on 
$\{1,2,\cdots,N-n\}$ and $\mathfrak{S}_{n}$
that acting on $\{N-n+1,N-n+2,\ldots,N\}$.
For $0\leq n\leq N$ and 
$R\in\mathfrak{S}_{N}/(\mathfrak{S}_{N-n}\times\mathfrak{S}_{n})$,
all the Bethe eigenstates $|k_{R};n\rangle$ with a set of momenta
$k_{R}$ have the same energy eigenvalue $E=\sum_{i=1}^{N}k_{i}$.
In the limit $L\to\infty$,
the Bethe eigenstates $|k_{R};n\rangle$ satisfy the relation
\begin{align}
 \langle k_{R};n|k_{S};m\rangle
&=\delta_{nm}\delta_{RS}L^{N}
  \prod_{i=1}^{N}\frac{2}{1+\cos\delta_{k_{i}}}+O(L^{N-1})
 \nn
\end{align}
for generic values of $\{k_{i}\}$.
Hence the normalized Bethe eigenstates are orthogonal 
in the limit $L\to\infty$.
As a result, the total degree of degeneracy of 
the energy eigenvalue $E=\sum_{i=1}^{N}k_{i}$ 
is $\sum_{n=0}^{N}\,_{N}C_{n}=2^{N}$.

We obtain a general $N$-electron eigenstate
by taking a linear combination of the $2^{N}$ degenerate
Bethe eigenstates $|k_{R};n\rangle$ in the form
\begin{align}
\label{eq:N-state_IRLM_2}
&|k\rangle
 =\sum_{n=0}^{N}\sum_{R}
 \sgn(R)A^{(n)}_{R}|k_{R};n\rangle,
\end{align}
where the sum on $R$ runs over elements in 
$\mathfrak{S}_{N}/(\mathfrak{S}_{N-n}\times\mathfrak{S}_{n})$.
The square norm of the eigenstate $|k\rangle$
is readily calculated from
$\langle k|k\rangle=\sum_{n=0}^{N}\sum_{R}|A^{(n)}_{R}|^{2}
\langle k_{R};n|k_{R};n\rangle$.

The expectation value
$\langle I\rangle=\langle k|I|k\rangle/\langle k|k\rangle$
of the current operator $I$ in \eqref{eq:current-op}
for each eigenstate 
$|k\rangle$ in \eqref{eq:N-state_IRLM_2}
is exactly given by
\begin{align}
\label{eq:current}
\langle I\rangle
&=\frac{-t}{2^{N-1}L}
 \sum_{n=1}^{N}\frac{n}{(N\!-\!n)! n!}
 \sum_{P\in\mathfrak{S}_{N}}\hspace{-5pt}
 \mathrm{Im}
 \big(A^{(n)\ast}_{P}A^{(n-1)}_{P}
 \e^{\frac{\i}{2}\delta_{k_{P_{N-n+1}}}}\big)
 \nn\\
&\quad\times\Big(
 \!\prod_{i=1}^{N-n}\hspace{-2pt}
 \cos\frac{\varphi_{k_{P_{i}}k_{P_{N-n+1}}}
 \!\!\!\!+\!\eta}{2}
 \Big)
 \Big(\!
 \cos\frac{\eta}{2}
 \Big)^{\!\!n-1}
 \cos\frac{\delta_{k_{P_{N-n+1}}}}{2}
 e_{k_{P_{N-n+1}}}
 +O(L^{-2}).
\end{align}
Here, by using the fact that any element $P\in\mathfrak{S}_{N}$ 
is decomposed as $P=RQ$ with a unique element $R\in\mathfrak{S}_{N}/%
(\mathfrak{S}_{N-n}\times\mathfrak{S}_{n})$ 
and a unique element $Q\in\mathfrak{S}_{N-n}\times\mathfrak{S}_{n}$,
we set $A^{(n)}_{P}=A^{(n)}_{R}$ for every $P$ with the same $R$.

The result in~\eqref{eq:current} 
shows that we need to consider the linear combination 
\eqref{eq:N-state_IRLM_2} of the degenerate Bethe states
in order to obtain a non-zero current expectation value.
Indeed, a specific Bethe eigenstate 
$|k_{R};n\rangle$ gives the expectation value
$\langle k_{R};n|I|k_{R};n\rangle=0$.
We stress that we do not impose periodic boundary conditions
to the eigenfunctions $g^{(n)}_{k}(x;y)$ and $e^{(n)}_{k}(x;y)$
in~\eqref{eq:N-Bethe-eigenfunction};
with the periodic boundary 
conditions~\cite{Konik-Saleur-Ludwig_01PRL,%
Konik-Saleur-Ludwig_02PRB}, 
the eigenstates $|k_{R};n\rangle$ with
different $n$ or different $R$ would not be degenerate,
and hence the current expectation value would be zero.

By expressing the eigenstate $|k\rangle$
in terms of the leads 1 and 2,
the eigenfunction describing $N-n$ electrons 
in the lead 1 and $n$ electrons in the lead 2 is given by
\begin{align}
(N\!-\!n)!n!\,F^{(n)}_{k}(z)
&\Define\langle c_{2}(z_{N})\cdots c_{2}(z_{N-n+1})
 c_{1}(z_{N-n})\cdots c_{1}(z_{1})|k\rangle
 \nn\\
&=\frac{1}{2^{\frac{N}{2}}}
  \sum_{m=0}^{N}
  \sum_{R}
  (-)^{\sharp\{R_{i}|N-n<R_{N-m+1},\ldots,R_{N}\}}
  \sgn(R)
 \nn\\
&\quad\times
  \langle c_{o}(z_{R_{N}})\cdots c_{o}(z_{R_{N-m+1}})
 c_{e}(z_{R_{N-m}})\cdots c_{e}(z_{R_{1}})|k\rangle,
 \nn
\end{align}
where $\sharp A$ stands for 
the number of elements in the set $A$.
We consider the behavior of $F^{(n)}_{k}(z)$ 
in the region $z_{1}<z_{2}<\cdots<z_{N}<0$.
The eigenfunction $F^{(n)}_{k}(z)$ is 
a complicated linear combination of
plane waves $\e^{\i\sum_{i}k_{P_{i}}z_{i}}$
for $P\!\in\!\mathfrak{S}_{N}$.
Among them, we call the plain wave 
$\e^{\i\sum_{i}k_{i}z_{i}}$ an ``incoming wave''.
The terms with the incoming wave are summarized as
\[
 \e^{\i\sum_{i}k_{i}z_{i}}\!\!
 \sum_{m, R}
  (-)^{\sharp\{R_{i}|N-n<R_{N\!-\!m\!+\!1},\ldots,R_{N}\}}
  \Tilde{A}_{R}^{(m)}
\]
where 
\[
 \Tilde{A}_{R}^{(m)}=A_{R}^{(m)}
 \e^{-\frac{\i}{2}(
 \sum_{i<j\leq N\!-\!m}
 \varphi_{k_{R_{i}}\!k_{R_{j}}}\!+
 \eta\sum_{i\leq N\!-\!m\atop N\!-\!m<j}
 \sgn(R_{i}\!-R_{j}))}.
\]

We {\it define} the scattering states $|k\rangle^{(\ell)}$,
($\ell=0,1,\ldots,N$)
by taking the coefficients $\{\Tilde{A}_{R}^{(n)}\}$ of
the eigenstate $|k\rangle$ as
\begin{align}
\label{eq:scattering}
 \Tilde{A}_{R}^{(n)}
&=(-)^{\sharp\{R_{i}|N-\ell<R_{N-n+1},\ldots,R_{N}\}},
\end{align}
for $k_{1}>k_{2}>\cdots>k_{N}$.
In the scattering state $|k\rangle^{(\ell)}$,
the incoming wave $\e^{\i\sum_{i=1}^{N}k_{i}z_{i}}$
exists only in the eigenfunction $F^{(\ell)}_{k}(z)$,
which describes $N-\ell$ electrons in the lead 1
and $\ell$ electrons in the lead 2.

The scattering states $|k\rangle^{(\ell)}$
are different from those 
of the standard one-body scattering theory in quantum mechanics.
If we were solving the one-body scattering problem, 
the scattering state would be obtained from the condition that
an electron comes only from the lead 1 or the lead 2.
However, the eigenstate
$|k\rangle$ in~\eqref{eq:N-state_IRLM_2} does not give 
such scattering state.
In fact, the scattering state $|k\rangle^{(\ell)}$
extends to all parts of the two leads for $U>0$.
In other words, it is impossible to judge whether each electron
comes from the lead 1 or the lead 2 for $U>0$,
which is not strange since we assume the same Fermi energy 
for both leads.
In the limit $U\to 0$, our scattering state $|k\rangle^{(\ell)}$
is reduced to the standard one-body scattering state.

\begin{figure}
\includegraphics[width=73mm,clip]{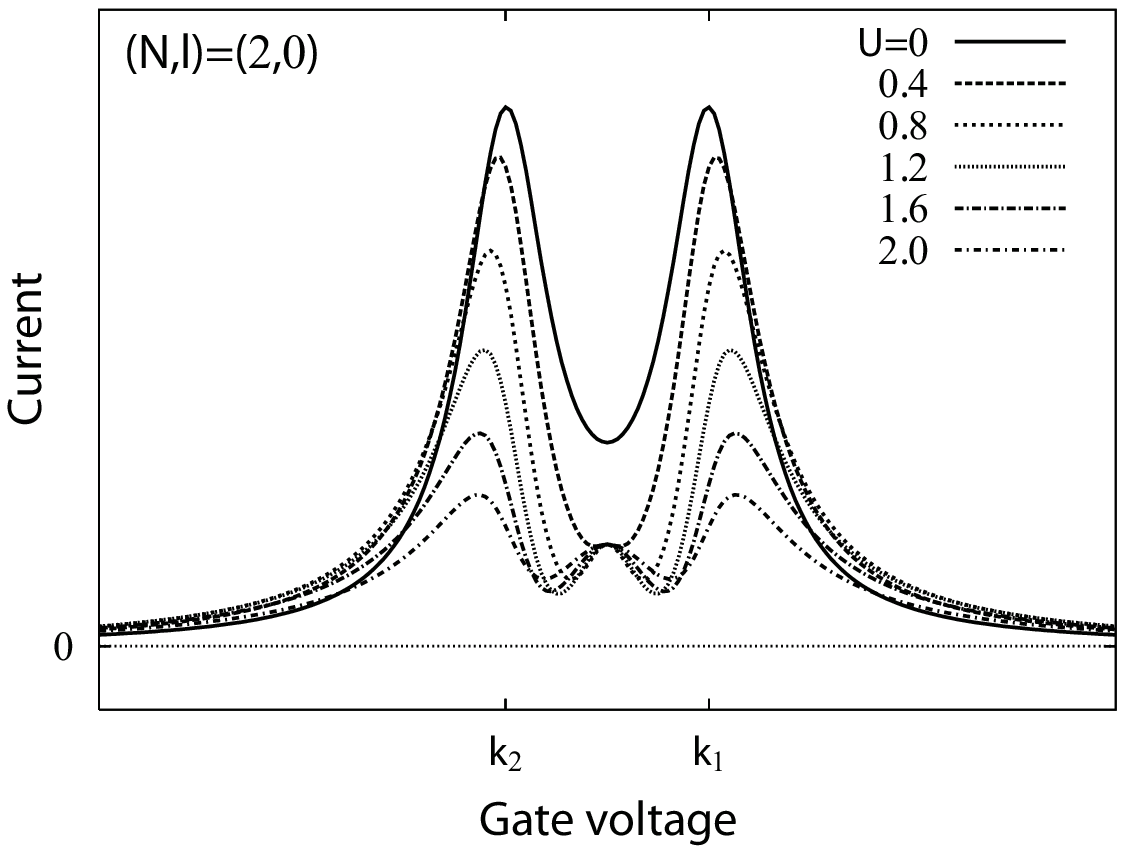}
\quad
\includegraphics[width=74mm,clip]{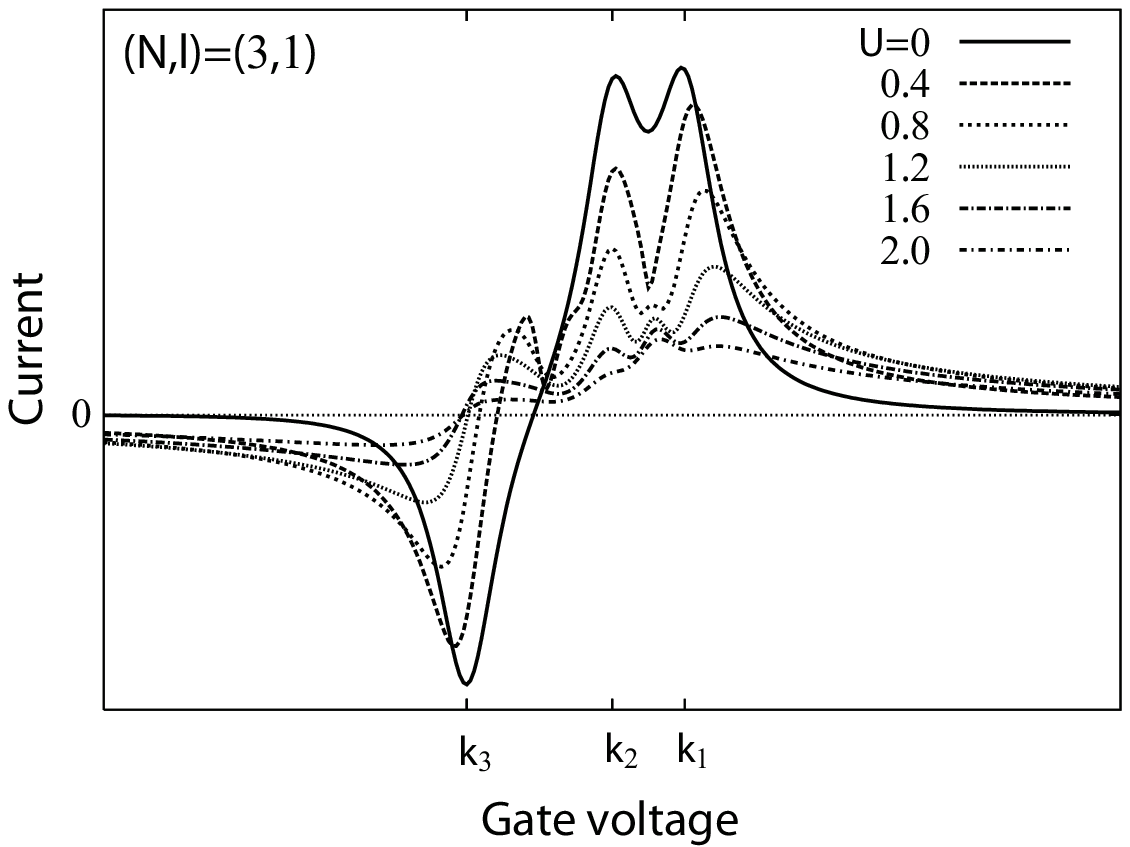}
\caption{\label{fig:resonance} 
The current expectation value $\langle I\rangle^{(\ell)}$
for the scattering states. We fixed $t=1$.}
\end{figure}

By applying \eqref{eq:scattering} 
to the expectation value \eqref{eq:current}, we have 
\begin{align}
&\langle I\rangle^{(\ell)}=
 \frac{t}{2^{N-1}L}
 \sum_{n=1}^{N}\frac{n}{(N\!-\!n)! n!}
 \sum_{P\in\mathfrak{S}_{N}}\!
 \sgn(P_{n}\!-\! N\!+\!\ell)
 \nn\\
&\times\!
 \sin\Big(\frac{1}{2}\Big(
 \sum_{i=1}^{n-1}
  \sgn(P_{i}\!-\!P_{n})
  \varphi_{k_{P_{i}}k_{P_{n}}}
 +\sum_{i=1}^{N}
  \sgn(P_{i}\!-\!P_{n})\eta
 -\delta_{k_{P_{n}}}
 \Big)\Big)
 \nn\\[-5pt]
&\times\!
 \Big(\!\prod_{i=1}^{n-1}\!
 \cos\frac{\varphi_{k_{P_{i}}k_{P_{n}}}\!\!\!\!+\eta}{2}
 \Big)
 \Big(\!
 \cos\frac{\eta}{2}
 \Big)^{\!N-n}\hspace{-10pt}
 \cos\frac{\delta_{k_{P_{n}}}}{2}e_{k_{P_{n}}}\!+\!O(L^{-2}).
 \nn
\end{align}
Short calculations reveal that
every term in $\langle I\rangle^{(\ell)}$
contains the product of the factors
$\sin\delta_{k},1+\cos\delta_{k},
\sin(\varphi_{k_{i}k_{j}}+\eta)$ and
$1+\cos(\varphi_{k_{i}k_{j}}+\eta)$,
which are rational functions of
$k_{i}, t, \epsilon_{d}$ and $U$.
The factors have poles
at $\epsilon_{d}=k_{i}\pm\i t^{2}/2,
(k_{i}+k_{j})/2\mp\i(k_{i}-k_{j})U/4$ 
in the complex plane of $\epsilon_{d}$.
Figure \ref{fig:resonance} shows the current expectation value 
$\langle I\rangle^{(\ell)}$ as a function 
of the gate energy $\epsilon_{d}$
for the scattering states 
$|k\rangle^{(\ell)}$ indexed by $(N,\ell)=(2,0)$ and $(3,1)$.

We find resonance peaks in the vicinity
of $\epsilon_{d}=(k_{i}+k_{j})/2$,
which correspond to many-body scattering;
they appear only for $U>0$.
As is stressed above,
the resonance of many-body scattering is originated from
the phase shifts which are different
for each two-body scattering in the lead $e$, in the lead $o$ 
and between the two leads.
We also find resonance peaks in the vicinity 
of $\epsilon_{d}=k_{i}$,
which correspond to one-body scattering at the quantum dot and 
are reduced to Lorentzian peaks in the limit $U\to 0$.
The resonance peaks in the vicinity of 
$\epsilon_{d}=(k_{i}+k_{j})/2$ were not present in 
Mehta and Andrei's result~\cite{Mehta-Andrei_06PRL};
their results are equal to the limit $U\to 0$ of our result.
This is because the interaction effect would be canceled
in the current expectation value $\langle I\rangle$ 
if we adopted the same phase shifts 
for all the two-body scattering in the lead $e$, in the lead $o$
and between the two leads.
Our choice \eqref{eq:2-Bethe-eigenfunction}
of the phase shifts of two-body scattering
is more plausible in the context of eigenstates
as mentioned above.
It would be interesting to discuss how the resonance of
many-body scattering
affects the transport properties of the 
interacting open quantum system out of equilibrium.

\section*{Acknowledgments}

The authors would like to thank Prof.~T.~Deguchi,
Dr.~T.~Imamura, Dr.~K.~Sasada and Dr.~M.~Matsuo 
for helpful comments.
The present study is partially supported by 
Core Research for Evolutional Science and Technology 
of Japan Science and Technology Agency.



\begin{thebibliography}{99} 
\bibitem{Andrei_80PRL}
N.~Andrei, 
Phys.~Rev.~Lett. \textbf{45} (1980), 379.

\bibitem{Andrei-Furuya-Lowenstein_83RMP}
N.~Andrei, K.~Furuya and J.~H.~Lowenstein, 
Rev.~Mod.~Phys. \textbf{55} (1983), 331.

\bibitem{Filyov-Wiegmann_80PLA}
V.~M.~Filyov and P.~B.~Wiegmann, 
Phys.~Lett.~A \textbf{76} (1980), 283.

\bibitem{Tsvelick-Wiegmann_83AP}
A.~M.~Tsvelick and P.~B.~Wiegmann, 
Adv.~Phys. \textbf{32} (1983), 453.

\bibitem{Wiegmann_80PLA}
P.~B.~Wiegmann, 
Phys.~Let.~A \textbf{80} (1980), 163.

\bibitem{Konik-Saleur-Ludwig_02PRB}
R.~M.~Konik, H.~Saleur and A.~Ludwig, 
Phys.~Rev.~B \textbf{66} (2002), 125304.

\bibitem{Konik-Saleur-Ludwig_01PRL}
R.~M.~Konik, H.~Saleur and A.~W.~W.~Ludwig, 
Phys.~Rev.~Lett. \textbf{87} (2001), 236801.

\bibitem{Mehta-Andrei_06PRL}
P.~Mehta and N.~Andrei, 
Phys.~Rev.~Lett. \textbf{96} (2006), 216802.

\bibitem{Cronenwett_98Science}
S.~M.~Cronenwett, T.~H.~Oosterkamp and L.~P.~Kouwenhoven, 
Science \textbf{281} (1998), 540.

\bibitem{GoldhaberGordon_98PRL}
D.~Goldhaber-Gordon, J.~Gores, M.~A.~Kastner,
H.~Shtrikman, D.~Mahalu and U.~Meirav, 
Phys.~Rev.~Lett. \textbf{81} (1998), 5225.

\bibitem{GoldhaberGordon_98Nature}
D.~Goldhaber-Gordon, H.~Shtrikman, D.~Mahalu,
D.~Abusch-Magder, U.~Meirav and M.~A.~Kastner, 
Nature (London) \textbf{391} (1998), 156.

\bibitem{Ralph-Buhrman_92PRL}
D.~C.~Ralph and R.~A.~Buhrman, 
Phys.~Rev.~Lett. \textbf{69} (1992), 2118.

\bibitem{Ralph-Buhrman_94PRL}
D.~C.~Ralph and R.~A.~Buhrman, 
Phys.~Rev.~Lett. \textbf{72} (1994), 3401.

\bibitem{Meir-Wingreen-Lee_91PRL}
Y.~Meir, N.~S.~Wingreen and P.~A.~Lee, 
Phys.~Rev.~Lett. \textbf{66} (1991), 3048.

\bibitem{Yeyati-MartinRodero-Flores_93PRL}
A.~Levy Yeyati, A.~Martin-Rodero and F.~Flores, 
Phys.~Rev.~Lett. \textbf{71} (1993), 2991.

\bibitem{Datta}
S.~Datta,
Electronic Transport in Mesoscopic Systems 
(Cambridge University, Cambridge, England, 1995).

\bibitem{Zagoskin}
A.~M.~Zagoskin,
Quantum Theory of Many-Body Systems, 
Graduate Texts in Contemporary Physics (Springer, New York, 1998).

\bibitem{Schiller-Hershfield}
A.~Schiller and S.~Hershfield, 
Phys.~Rev.~B \textbf{58} (1998), 14978.
\end{thebibliography}
\end{document}